\documentstyle[12pt,moriond,psfig]{article}
\begin{document}
\heading{The CRONA (Caltech-Roma-Napoli) project}
\author{S. Andreon$^1$, S. Zaggia$^1$, R. de Carvalho$^2$, S.  
Djorgovski$^3$,\\G. Longo $^1$, I. Musella$^1$, R. Scaramella$^4$} 
{$^{1}$ Osservatorio Astronomico di Capodimonte, Naples, Italy} 
{$^{2}$ Observatorio Nacional de Rio de Janeiro, Brasil}
{$^{3}$ Palomar Observatory, Caltech, USA} 
{$^{4}$ Osservatorio Astronomico di Roma, Monteporzio, Italy} 
\centerline{\psfig{figure=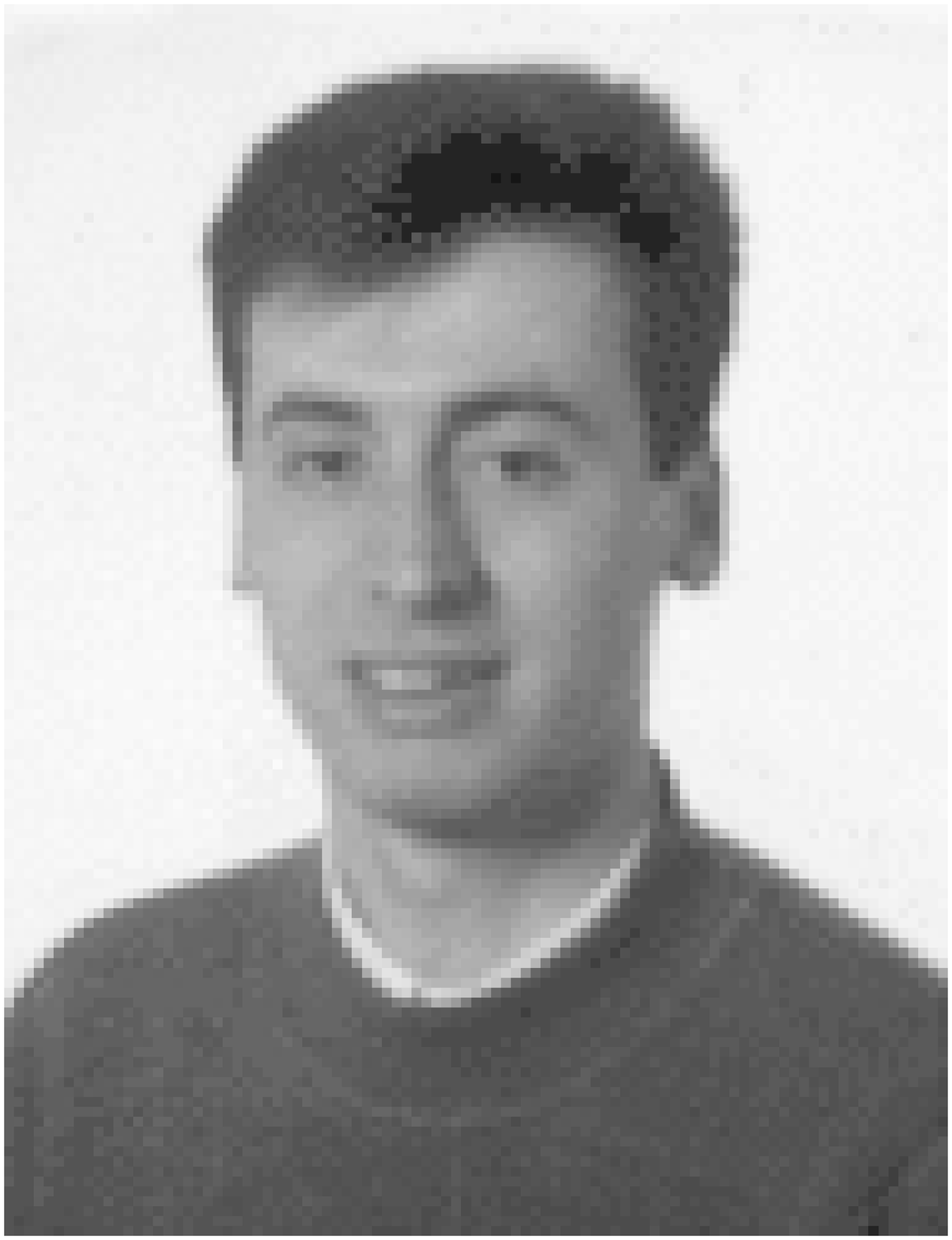,height=3.8truecm}}
\begin{moriondabstract}
The CRONA project is a joint enterprise between Caltech and the
astronomical observatory of Rome and Naples aimed to the photometric
calibration of the Second Palomar Sky Survey (POSS-II) plates and to the
construction of a complete photometric and astrometric catalogue of all
objects present on the POSS-II. The calibrated final catalogues of objects
will be made available to the community.  We present here 
a comparison between the performances of the SKICAT and DAOPHOT packages 
applied to POSS-II images of the globular cluster M92.
\end{moriondabstract}
\section{Introduction}

CRONA (Caltech-ROma-NApoli) is a recent extension of a lorg term project
started at Caltech in 1993 (Djorgovski et al. 1992) and aimed at the
construction of the so called Palomar Norris Sky Catalog (PNSC): i.e. a
complete photometric and astrometric catalogue of all objects present on
the Second Palomar Sky Survey (POSS-II) for an estimated number of $2
\times 10^9$ stars and $5 \times 10^7$ galaxies. 

The data processing pipeline consists of three steps: 
\begin{itemize}
\item 
digitalization of the plates, performed at ST-ScI with 15 $\mu$ (=1
arcsec) pixels. STScI provides also the astrometric plate solution (within
0.5 arcsec accuracy); 
\item 
plate processing and catalogs construction, performed with the SKICAT
software (Weir et al. 1995a) developed at Caltech; 
\item 
photometric calibration of the plates through the aquisition of CCD frames
of galaxy rich fields. 
\end{itemize}

Details on the expected scientific results and aims are given in 
Djorgovski et al. (1997); in this paper we shall focus on 
some aspects of the data quality and on some specific 
applications.

\setbox111=\hbox{$\vcenter{\psfig{figure=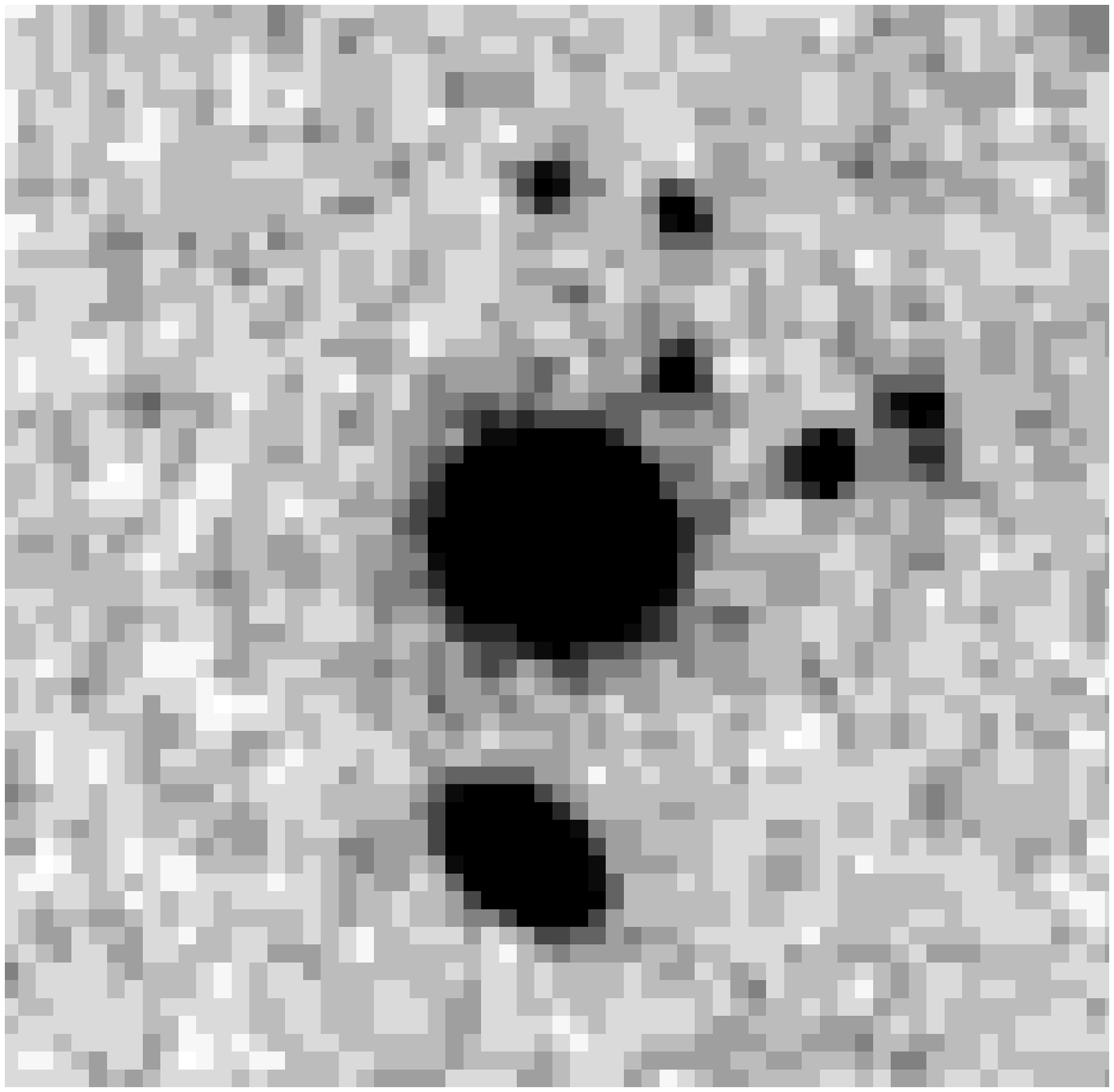,width=7truecm}}$}
\setbox112=\hbox{$\vcenter{\psfig{figure=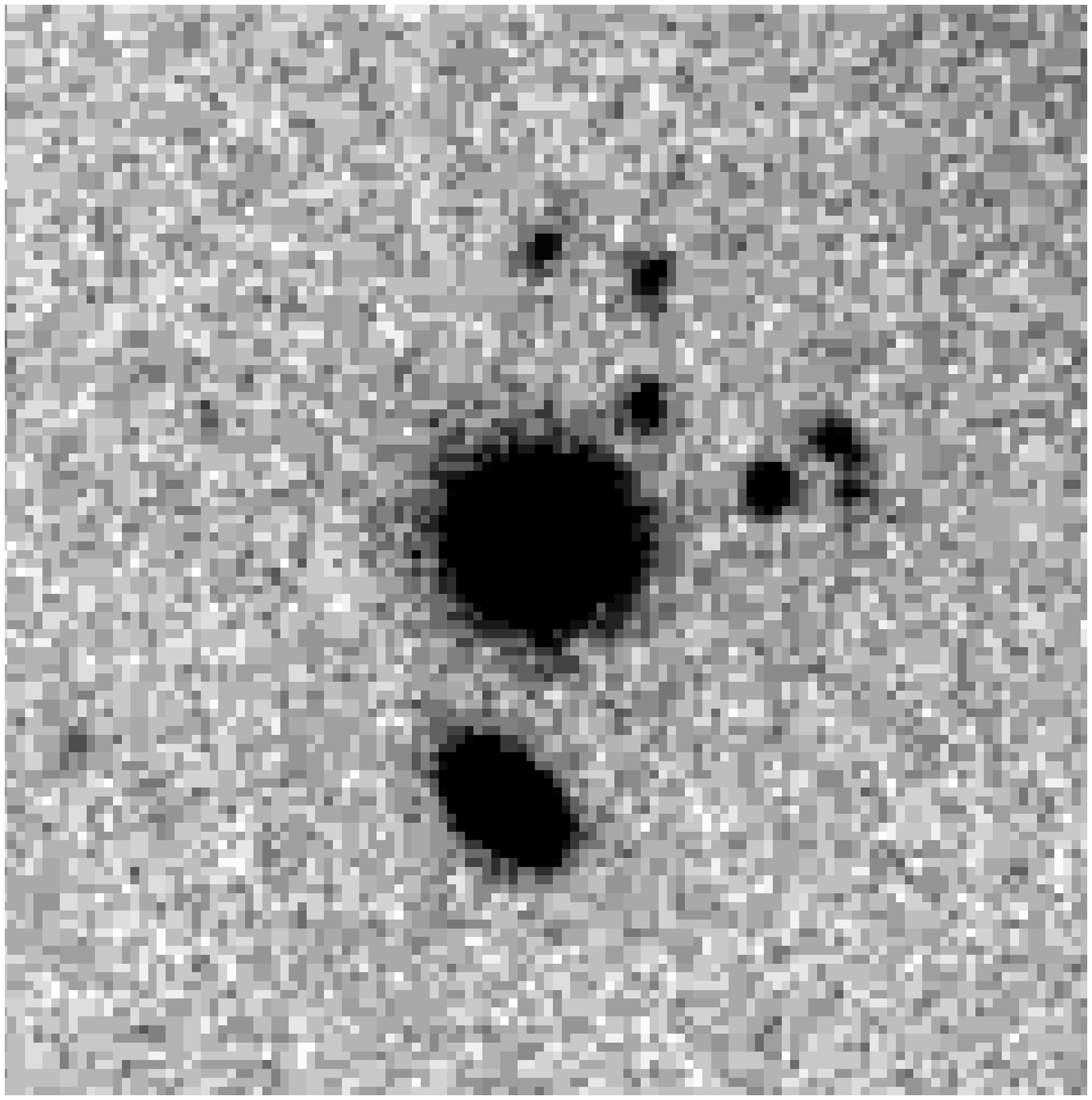,width=8.25truecm}}$}

\begin{figure}
\noindent
\centerline{\box111 \box112}
 
\noindent 
\caption{Comparison between POSS-I (left) and POSS-II (right) images
of galaxies in the Coma cluster direction}
\end{figure}

\section{The data}

POSS-II (Reid et al. 1991) covers all the Northern Sky (including the
celestial Equator) and consists of 894 plates (each $6.5 \times 6.5$ sq.
deg.) in three colors ($B_J$, $R_F$ and $I_N$) spaced by 5 degree on the
sky, i.e. with an overlap ($0.75$ deg. on each side) between adiacent
fields, much larger than for the POSS-I plates (which were taken only in
two colors and with a spacing between plate centers of about 6 deg). The
large overlap regions ensure several advantages: 40 \% of the sky is
registered at least twice in three colors; better photometric matching
between adiacent plates can be obtained, and most of the sky is imaged in the
central unvignetted part of the plates. Due to the improved sensitivity
and resolution of the emulsion, to the use of hypersensitization and long
exposure times, magnitude limits of the POSS-II material are typically
22.5, 20.8 and 19.5 in the $B_J$, $R_F$ and $I_N$ respectively (Reid et
al. 1991); i.e. about 1 mag fainter than for the POSS-I plates.
Furthermore POSS-II plates are provided with calibration spots (missing on
POSS-I), thus allowing density-intensity conversion of the data. 
Figure 1 gives a visual impression of the difference between POSS-I
and POSS-II images for the same Coma cluster field.

Absolute photometric calibration of the plate material is obtained through
CCD photometry of at least two fields per plate, taken in fair seeing
conditions (better than 2 arcsec) in the $g$, $r$ and $i$ bands of the
Thuan--Gunn photometric system (Thuan \& Gunn 1976, Kent 1985). Being
CRONA mainly aimed to the study of the extragalactic universe, 
CCD fields are selected to contain a fairly large number of galaxies, in
order to attain more accurate photometric calibration at low light levels,
and to minimize color mismatch and reciprocity effects. Selection of
target fields is performed either from the Abell and Zwicky catalogues
(Abell 1958, Zwicky 1961-1968) and/or from the CRONA catalogues
themselves. 

There is some mismatch between the plate and CCD photometric systems: $F$
and $N$ response plates are reasonably matched by the $r$ an $i$, while
the $J$ is larger and bluer than the $g$ band and some color term
correction needs to be taken into account. So far, CCD observations are
made at the 1.5 m Palomar and Loiano (Bologna) telescopes, at ESO (an
equatorial zone, $\delta<10$ degrees, 10 nights allocated) and soon at the
new 1.5m TT1 telescope in Southern Italy. The achieved photometric accuracy,
both across and within the plates and including both photometric and systematic 
errors, ranges from 0.05 mag at the bright end to 0.2 at $r \sim 21.5$ for 
stellar objects (Djorgovski et al. 1997).  For extended objects,
photometric accuracy is typically 50 \% worse than for stellar ones.
Such low values result from both the proper managing of the plate data and
from the strong requirements on the quality of CCD calibration images. 

Plates are currently being reduced at three center, Caltech, Naples and Rome, 
with identical machines and software (SKICAT). After the arrival of one of us 
(RdC) to the Observatorio Nacional in Rio de Janeiro, also this institute will 
soon join in. 

Due to vignetting, non linearity of the plate response, position
dependence of the point spread function and large space disk occupation of
the data (1 GB per plate), plate reduction is a tricky business. For
detailed discussions, see Fayyad et al. (1996) and Weir et al. (1995a and
b). Few items need however to be stressed here.  The object detection algorithm
in SKICAT is an improved version of FOCAS (Jarvis \& Tyson 1979) which
allows the use of a fixed flux detection threshold for fields in which the
noise is not uniform.  Star/galaxy classification is performed from the
catalogues via decision trees trained on the CCD calibration
frames (Weir et al. 1995a).  In order to take into account the variation of
the PSF shape as a function of the position on the plate, the object classification is
performed locally - id est in each ''small'' portion of the plate. At
20 mag in $r_F$ the galaxy catalogue is 90 \% complete with a 20\%
contamination from stars (Weir et al 1995b). 

The true bottle neck of the project is in the allocation of
telescope time for the aquisition of the CCD calibration frames. 
It is reasonable to expect, however, that the whole set of plates 
will be calibrated before year 2000. 

The main goals of the CRONA collaboration are, besides the completion of
the PNSC itself, i) the study of large scale structure via the compilation
of an objective catalogue of putative galaxy clusters;  ii) the derivation
of accurate lumnosity functions for a large sample of galaxy clusters; 
iii) the search for high redshift QSO's (in the range of $z$ 3.8-4.3,
Kennefick et al. 1995); iv) cross correlation of the CRONA catalogues with
catalogues obtained at other wavelenghts. 
A variety of interesting results can however be obtained also in other 
fields of astrophysical research. 

\setbox113=\hbox{\psfig{figure=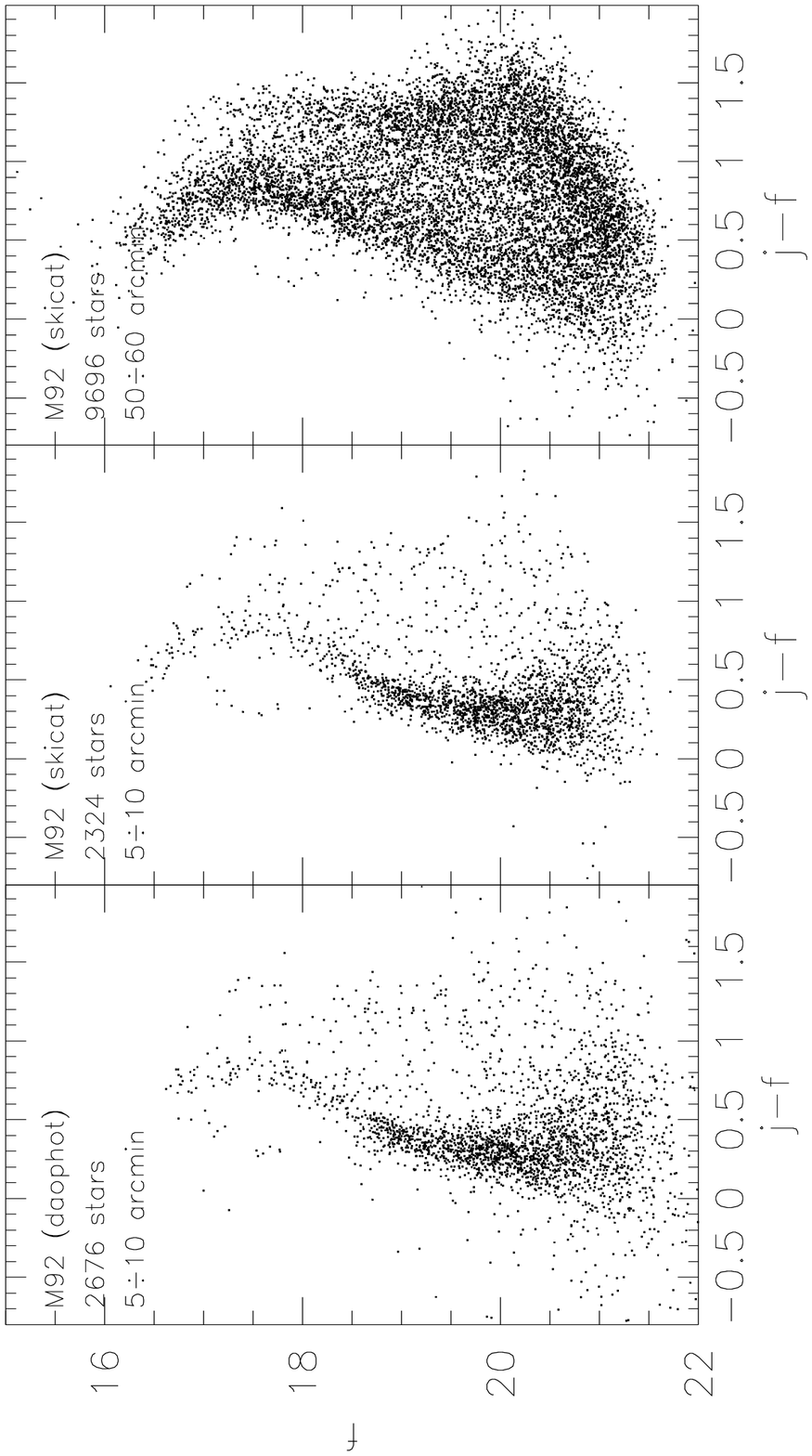,height=24truecm}}

\begin{figure}
\noindent
\centerline{\box113}
 
\noindent 
\caption{Left and central panel: comparison between the ($F$, $J-F$) CMD 
obtained with DAOPHOT (left) and with SKICAT (center) in the radial 
range $5\div10$ arcmin. The right panel shows the CMD of a radial anulus 
located at $50\div60$ arcmin from the center of the cluster. 
Magnitudes are instrumental (i.e. not calibrated).}
\end{figure}

\section{Color--magnitude diagram of M92}

Some further indications on the quality of the POSS-II data can be
inferred from the study of the color--magnitude diagram (CMD) of the
globular cluster M92 (NGC 6341). In Fig.~2 we show the uncalibrated ($F,
J-F$) diagram for the cluster M92 obtained from the $J$ and $F$ plates of
the POSS-II field n. 278. The left and the central panels refer to an
annular region comprised between 5 and 10 arcmin from the center of the
cluster (the region below 5 arcmin is completely overexposed and
overcrowded and no information can be extracted). The central and right
panels show the CMD build from the SKICAT core magnitudes, 
defined as aperture photometry in a square box of $3''\times3''$. 
In order to evaluate the quality of the
POSS-II data and of the SKICAT outputs we also derived a CMD of M92 using 
DAOPHOT (Stetson 1987) aperture photometry. The left panel
in Fig.~2 gives the DAOPHOT CMD in the same radial range of the central
panel ($5\div10$ arcmin).  The DAOPHOT CMD required an interactive 
definition of the PSF for each plate after that the plates had been 
linearized using the density--intensity calibration curve provided by SKICAT.
The positions of the objects fitted by ALLSTAR were then used to derive
aperture magnitudes for each star. Since DAOPHOT calculates aperture
magnitudes in circular spot, we used a circle having the same area of the
square box used by SKICAT in deriving core magnitudes. 

When comparing the left and central panel of Fig.~2, few points are worth
noticing: first, the two CMDs appears to be almost similar with a scatter
of the data 1 mag below the turnoff point (at $F\sim20.5$) identical for the
two plots; second, there are more objects in the DAOPHOT plot; third, the
SKICAT CMD appears to be {\it cleaner\/} than the DAOPHOT one. We have to
stress that DAOPHOT magnitudes have been obtained after some interactive
work and some fine tuning on images already linearized by SKICAT, whereas
SKICAT magnitudes come from the stardard output of the automatic procedure. 

The larger number of objects detected by DAOPHOT is mainly due to:
contamination from galaxies (which are {\it automatically} removed by
SKICAT), mismatching of stars (SKICAT matchs the plates using the ST-ScI
astrometric solution of the plates whereas DAOPHOT data were matched in
pixel space), a deeper detection threshold used for DAOPHOT, and
a slightly better effectiveness of DAOPHOT in dealing with crowded fields.
The cleaner apparence of the SKICAT CMD derives, instead, mainly by 
the lower contamination by mismatching and by the rejection of
non-stellar object (galaxies, plate scratches, etc.).

The right panel of Fig.~2 shows the CMD for an annular region centered on
the cluster and comprised between 50 and 60 arcmin.  At such large
distances from the center of the cluster, most objects are likely
galactic stars with only a small fraction belonging to the cluster. 
The large statistics allows an accurate evaluation of the background and,
therefore, permits the derivation of star counts up to large clustercentric radii
as well as higher accuracy in deriving the outer isoplets and the tidal radius of the
cluster. 

A second hint on the quality of the data comes from the thickness
of the cluster MS. Accordingly to Renzini and Fusi Pecci (1988) the
intrinsic thickness of the MS is expected to be negligible and therefore the
observed thickness ($\simeq0.16$~mag at $F=20.5$, i.e. $\sim 1$ mag below
the turnoff point) may be attributed to galactic contamination and
observational errors only. 
A third and last hint on the quality of the data comes from the comparison
of the SKICAT CMD with those derived by other authors 
who have performed similar work.
The SKICAT CMD of M92 is deeper and looks better defined and thinner than those 
derived for other nearby globular clusters by Grillmair et al. (1995) or by Lehmann 
and Scholz (1997) who used UKST and Tautenburg plates scanned with APM. 

As it was to be expected the SKICAT CMD looks noiser and less deep than 
that of M\,55, measured by Zaggia et al. (1997) who made use of a mosaic 
of CCD frames taken at the 3.5m ESO-NTT. These authors, however, 
could cover only a quadrant of the cluster and push their data only out to 
1.5 times its tidal radius without reaching the background. This undermines
the detection of those tidal tails and diffuse extra-tidal stars which were
first discovered, for other clusters, by Grillmair et al. (1995).

POSS-II data extracted by SKICAT allow to circumvent the present CCD field of view
limitations and a systematic study of the outer haloes of all globular clusters
present on the POSS-II plates is under progress.

\end{document}